\documentclass[10pt,twocolumn,prl,aps,superscriptaddress,floatfix,nobalancelastpage,
preprintnumbers,reprint,footinbib,hyperref]{revtex4-2}

\usepackage{graphicx}
\usepackage{dcolumn}
\usepackage{bm}
\usepackage[normalem]{ulem}
\usepackage{esvect}
\usepackage{amsmath}

\usepackage{ulem}

\usepackage{xcolor}
\definecolor{ochre}{rgb}{0.8, 0.47, 0.13}

\newcommand{\kicksOneSigma}{10}

\newcommand{\massOneSigma}{1.0}

\newcommand{\nuKicksOneSigma}{8}

\newcommand{\nuKickOneSigmaMax}{60}
\newcommand{\CoMKicksOneSigma}{17}

\newcommand{\peakMass}{0.3}
\newcommand{\peakVelocity}{4}
\newcommand{\peakVelocityNeutrino}{3}
\newcommand{\peakSystemicVelocity}{3}
\newcommand{\minPeriod}{5}
\newcommand{\maxPeriod}{15}
\newcommand{\minKick}{0}
\newcommand{\maxKick}{200}
\newcommand{\minMass}{0}
\newcommand{\maxMass}{5}
\newcommand{\maxAsymmetryFactor}{0.038}
\newcommand{\maxAsymmetryPerCent}{4}
\newcommand{\likelyAsymmetryPerCent}{0.2}
\newcommand{\minAsymmetryFactorLimForMaxLikelihood}{2\times10^{-4}}
\newcommand{\maxAsymmetryFactorLimForMaxLikelihood}{2\times10^{-3}}
\newcommand{\lastAsymmetryFactorLimForMaxLikelihood}{5\times10^{-3}}

\def\aj{AJ}
\def\mnras{Mon. Not. R. Astron. Soc.}
\def\apjl{Astrophys. J. Lett.}
\def\apjs{Astrophys. J. Suppl. Ser.}
\def\pasa{Publications of the Astronomical Society of Australia}
\def\aap{Astron. Astrophys.}
\def\bain{Bulletin Astronomical Institute of the Netherlands}

\makeatletter

\newcommand{\checknextarg}{\@ifnextchar\bgroup{\gobblenextarg}{}}
\newcommand{\gobblenextarg}[1]{\,\mathrm{#1}\@ifnextchar\bgroup{\gobblenextarg}{}}
\makeatother

\usepackage{tikz,xcolor,hyperref}

\definecolor{lime}{HTML}{A6CE39}
\DeclareRobustCommand{\orcidicon}{\hspace{-1mm}
	\begin{tikzpicture}
	\draw[lime, fill=lime] (0,0)
	circle [radius=0.16]
	node[white] {{\fontfamily{qag}\selectfont \tiny \,ID}};
	\draw[white, fill=white] (-0.0525,0.095)
	circle [radius=0.007];
	\end{tikzpicture}
	\hspace{-3mm}
}

\foreach \x in {A, ..., Z}{\expandafter\xdef\csname orcid\x\endcsname{\noexpand\href{https://orcid.org/\csname orcidauthor\x\endcsname}
			{\noexpand\orcidicon}}
}

\begin{document}


\title{Constraints on neutrino natal kicks from  black-hole binary VFTS 243}

\author{Alejandro Vigna-G\'omez\orcidA{}}
\affiliation{%
Max-Planck-Institut f\"ur Astrophysik, Karl-Schwarzschild-Str.~1, 85748 Garching, Germany
}%
\affiliation{%
Niels Bohr International Academy, Niels Bohr Institute, Blegdamsvej 17, 2100 Copenhagen, Denmark
}%

\author{Reinhold Willcox\orcidB{}}
\affiliation{
 Institute of Astronomy, KU Leuven, Celestijnlaan 200D, 3001 Leuven, Belgium
}%
\affiliation{
 School of Physics and Astronomy, Monash University, Clayton, VIC 3800, Australia
}%
\affiliation{
 The ARC Centre of Excellence for Gravitational Wave Discovery -- OzGrav, Australia
}%

\author{Irene Tamborra\orcidC{}}
\affiliation{%
Niels Bohr International Academy, Niels Bohr Institute, Blegdamsvej 17, 2100 Copenhagen, Denmark
}%
\affiliation{%
DARK, Niels Bohr Institute, University of Copenhagen, Jagtvej 128, 2200 Copenhagen, Denmark
}%

\author{Ilya Mandel\orcidD{}}
\affiliation{
 School of Physics and Astronomy, Monash University, Clayton, VIC 3800, Australia
}%
\affiliation{
 The ARC Centre of Excellence for Gravitational Wave Discovery -- OzGrav, Australia
}%

\author{Mathieu Renzo\orcidE{}}
\affiliation{
Center for Computational Astrophysics, Flatiron Institute, 162 5th Ave, New York, NY 10010, USA
}
\affiliation{
Steward Observatory, University of Arizona, 933 N. Cherry Ave., Tucson, AZ 85721, USA
}%

\author{Tom Wagg\orcidF{}}
\affiliation{%
Department of Astronomy, University of Washington, Seattle, WA, 98195
}%

\author{Hans-Thomas Janka\orcidG{}}
\affiliation{%
Max-Planck-Institut f\"ur Astrophysik, Karl-Schwarzschild-Str.~1, 85748 Garching, Germany
}%

\author{Daniel Kresse\orcidH{}}
\affiliation{%
Max-Planck-Institut f\"ur Astrophysik, Karl-Schwarzschild-Str.~1, 85748 Garching, Germany
}%
\affiliation{%
Technical University of Munich, TUM School of Natural Sciences, Physics Department, James-Franck-Str. 1, 85748 Garching, Germany
}%

\author{Julia Bodensteiner\orcidI{}}
\affiliation{%
European Southern Observatory, Karl-Schwarzschild-Str. 2, 85748 Garching bei M\"unchen, Germany.
}%

\author{Tomer Shenar\orcidJ{}}
\affiliation{%
The School of Physics and Astronomy, Tel Aviv University, Tel Aviv 6997801, Israel
}%

\author{Thomas M. Tauris\orcidK{}}
\affiliation{%
Department of Materials and Production, Aalborg University, Denmark
}%


\date{\today}

\begin{abstract}
The recently reported observation of VFTS 243 is the first example of a massive black-hole binary system with negligible binary interaction following black-hole formation. The black-hole mass ($\approx 10\ M_{\odot}$) and near-circular orbit ($e\approx 0.02$)  of VFTS 243 suggest that the progenitor star experienced complete collapse, with energy-momentum being lost predominantly through neutrinos. VFTS 243 enables us to constrain the natal kick and neutrino-emission asymmetry during black-hole formation.  At $68\%$ C.L., the natal kick velocity (mass decrement) is $\lesssim \kicksOneSigma$ km/s ($\lesssim \massOneSigma\ M_\odot$), with a full probability distribution that peaks when $\approx \peakMass\ M_\odot$ were ejected, presumably in neutrinos, and the black hole experienced a natal kick of $\peakVelocity$ km/s. The neutrino-emission asymmetry is $\lesssim \maxAsymmetryPerCent\%$, with best fit values of $\sim0$--$\likelyAsymmetryPerCent\%$. Such a small neutrino natal kick accompanying black-hole formation is in agreement with theoretical predictions.
\end{abstract}
\maketitle

{\em Introduction.}---Stars several times more massive than the Sun end their lives with the collapse of their iron cores.
The explosion mechanism and the physics driving the formation of the compact object are still a matter of intense research~\cite{2012ARNPS..62..407J,2016ARNPS..66..341J,2016PASA...33...48M,2020LRCA....6....3M,2020LRCA....6....4M,2021Natur.589...29B,2022MNRAS.517..543W,2023IAUS..362..215M}. In the delayed neutrino-driven  mechanism, neutrinos revive the stalled shock wave, eventually leading to a successful explosion. In this case, the stellar mantle is successfully ejected and the compact-object remnant is a neutron star (NS) in most cases. However, if the explosion mechanism fails, continuous accretion of matter onto the transiently stable proto-NS pushes the latter over its mass limit and a black hole (BH) forms.

The extreme speeds of pulsars~\cite{2021RAA....21..141Y,2009ApJ...698..250C,2017A&A...608A..57V} indicate that NSs experience a natal kick during formation~\cite{2000ASSL..254..127L}. This natal kick is attributed mainly to asymmetric mass ejection, but asymmetric neutrino emission may contribute~\cite{1987IAUS..125..255W,1994A&A...290..496J,1996PhRvL..76..352B,2006A&A...457..963S,2012MNRAS.423.1805N,2013A&A...552A.126W, 2017ApJ...844...84H,2018ApJ...856...18K,2022MNRAS.517.3938C}. Simulations of the collapse of massive stars  show that hydrodynamical instabilities leading to large-scale asymmetries in the mass distribution---such as neutrino-driven convection~\cite{1994ApJ...435..339H,1995ApJ...450..830B,1996A&A...306..167J}, the standing accretion shock instability (SASI)~\cite{2003ApJ...584..971B,2006ApJ...642..401B} or the lepton emission self-sustained asymmetry (LESA) instability~\cite{2014ApJ...792...96T,2017MNRAS.471..914K}---could also have important consequences on  the natal kick of the compact remnant~\cite{2014ApJ...792...96T,2006A&A...457..963S,2012MNRAS.423.1805N,2013A&A...552A.126W}. In addition, fast rotation and the presence of magnetic fields at the moment of collapse are  likely to affect the natal kick \cite{2006A&A...457..963S}.

In the extreme scenario of \textit{complete collapse} into a BH, the ejecta mass and natal kicks are thought to be very low ($\sim$1-10 km/s) \cite{2013MNRAS.434.1355J,2020MNRAS.495.3751C,2022MNRAS.512.4503R}. In this case, mass-energy is lost via neutrinos and, to a lesser extent, gravitational waves \cite{2012ARNPS..62..407J,2022MNRAS.517.3938C}. This differs from the archetypical scenario in which anisotropic baryonic ejecta are the main carriers of momentum~\cite{2020MNRAS.496.2039S}.

Observations of stellar-mass BHs in high-mass X-ray binaries (HMXB) have been employed to constrain the impact of natal kicks of collapsing stars on the orbital configuration of massive binary systems \cite{1999MNRAS.310.1165T,2002ApJ...574..364P,2012ApJ...747..111W}. HMXBs are binary systems comprised of OB-type stars that transfer matter via stellar winds \footnote{In this \textit{Letter}, we do not consider the Be/X-ray binary class.} to their compact-object companion \cite{2006csxs.book..623T}. To date about $150$ HMXBs have been discovered in the Milky Way, but only a handful of these are associated with BH companions  \citep{2016A&A...587A..61C,2023A&A...671A.149F}.

\begin{figure}
    \centering
    \includegraphics[trim=0.7cm 0cm 0.5cm 0.5cm, clip,width=\columnwidth]{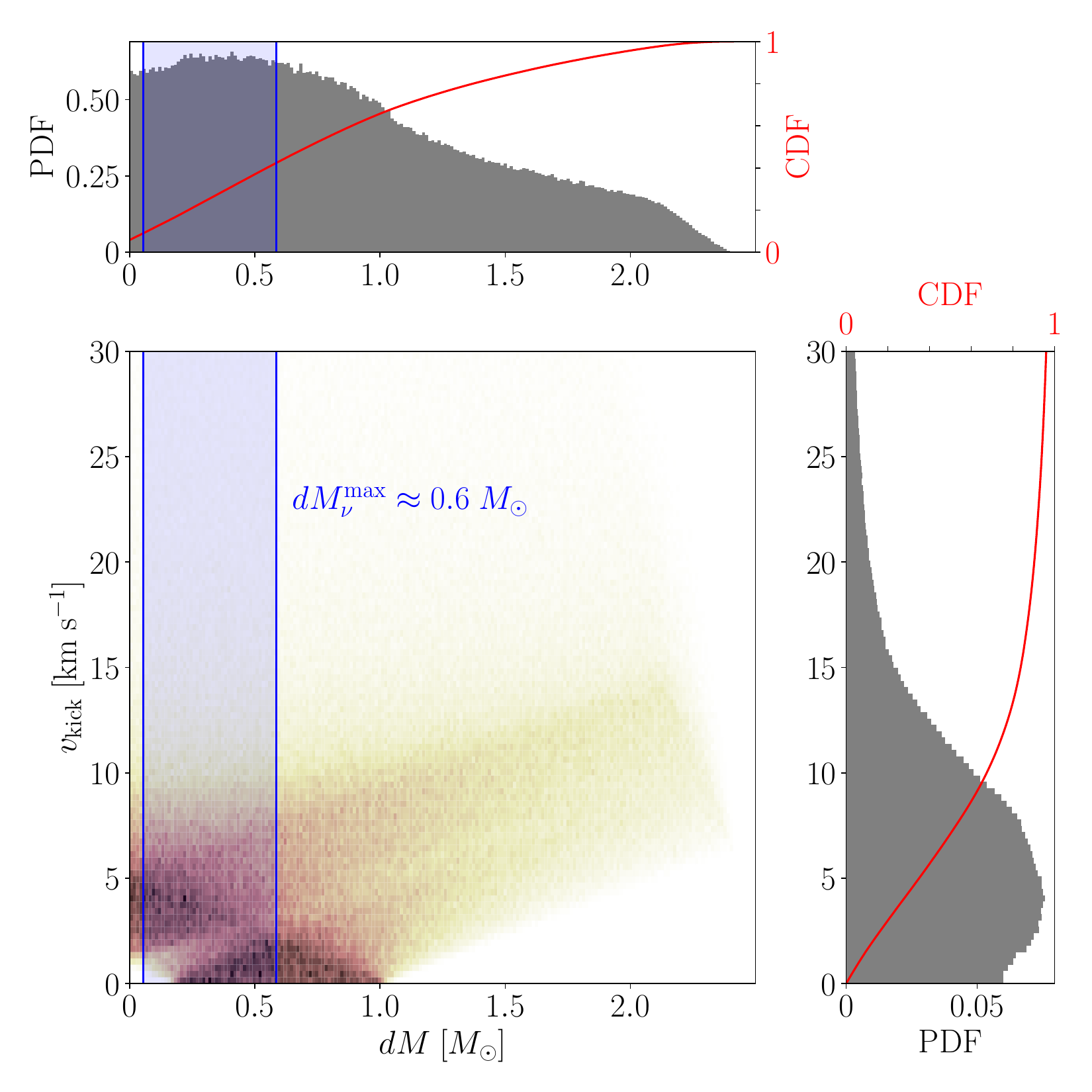}
    \caption{
    Constraints on the natal kick from VFTS 243. This figure shows the
    probability distribution of matching the current orbital period,
    eccentricity and systemic velocity of VFTS 243, given a certain mass
    loss ($dM$) and natal-kick magnitude ($v_{\rm{kick}}$) following
    BH formation. In the heatmap, darker (lighter) regions indicate a
    higher (lower) probability for a certain $dM$-$v_{\rm{kick}}$ pair
    to reproduce the  orbital configuration of VFTS 243. The blue
    shaded region, delimited by the solid blue vertical lines, shows
    the estimates for neutrino mass-loss decrements from stellar
    models of BH progenitors \cite{2021ApJ...909..169K}. The side
    panels display the marginalized probability distribution (PDF,
    gray histograms) and cumulative distribution functions (CDF, red
    curves). With
    $\rm{CDF}(\textit{dM}=\massOneSigma\ \textit{M}_{\odot})=\rm{CDF}( \textit{v}_{\rm{kick}}=\kicksOneSigma\ \rm{km/s})= 0.68$,
    we conclude that solutions with low mass loss and small natal
    kicks are preferred. The marginalized distributions peak at around
    $dM=\peakMass\ M_{\odot}$ and $v_{\rm{kick}}=\peakVelocity$ km/s,
    which is consistent with a solution where the mass decrement comes
    exclusively from neutrino emission. 
    }
    \label{fig:results}
\end{figure}

Recently, spectroscopy \cite{2021ApJ...913...48G,2022NatAs...6.1085S,2022A&A...665A.148S,2022A&A...664A.159M} and astrometry \cite{2023MNRAS.518.1057E,2023MNRAS.521.4323E} have allowed the first detection of \textit{inert} (i.e., X-ray quiescent or dormant) stellar-mass BH binaries \cite{2020A&A...638A..39L} in the field \footnote{In this \textit{Letter}, we will not consider observations of BH binaries from globular clusters \cite{2018MNRAS.475L..15G}, which are likely assembled dynamically}. This population of BH binaries is especially intriguing. These objects have wider orbital periods than HMXBs and the stellar companions are well within their Roche lobes. This configuration, which implies little interaction following BH formation (see Supplemental Material), makes inert BH binaries better probes of natal kicks than HMXBs. In this \textit{Letter}, we present direct inference on neutrino natal kicks for the most massive inert BH detected to date: VFTS 243 \cite{2022NatAs...6.1085S}. Our findings are summarized in Fig.~\ref{fig:results}.

For BH binaries, the plausibility of the complete collapse scenario can be assessed because the impact on the orbital evolution of mass ejection ($dM$) and a natal kick ($\bf{v}_{\rm{kick}}$) during compact object formation on binary star systems are well understood \cite{1995MNRAS.274..461B,1996ApJ...471..352K,1998A&A...330.1047T}. For example, near-instantaneous (quasi-)spherically symmetric mass ejection during the stellar collapse results in a recoil (the {Blaauw effect} \cite{1961BAN....15..265B}) that leads to a systemic velocity of the binary as a whole, widens the orbit, makes it more eccentric and can even disrupt the binary \cite{2019A&A...624A..66R}. On the other hand, asymmetries in the ejected mass lead to a wider variety of configurations, potentially modifying the separation (increase or decrease), eccentricity and inclination of the orbit. Such asymmetries could play an important role in the formation of BH-BH mergers \cite{2022ApJ...938...66T}, BH HMXBs \cite{1999A&A...352L..87N}, and inert BH binaries \cite{2023ApJ...959..106B,2023MNRAS.520.4740S}.

\begin{figure}
    \centering
    \includegraphics[trim=1cm 7.5cm 1cm 8.2cm, clip,width=\columnwidth]{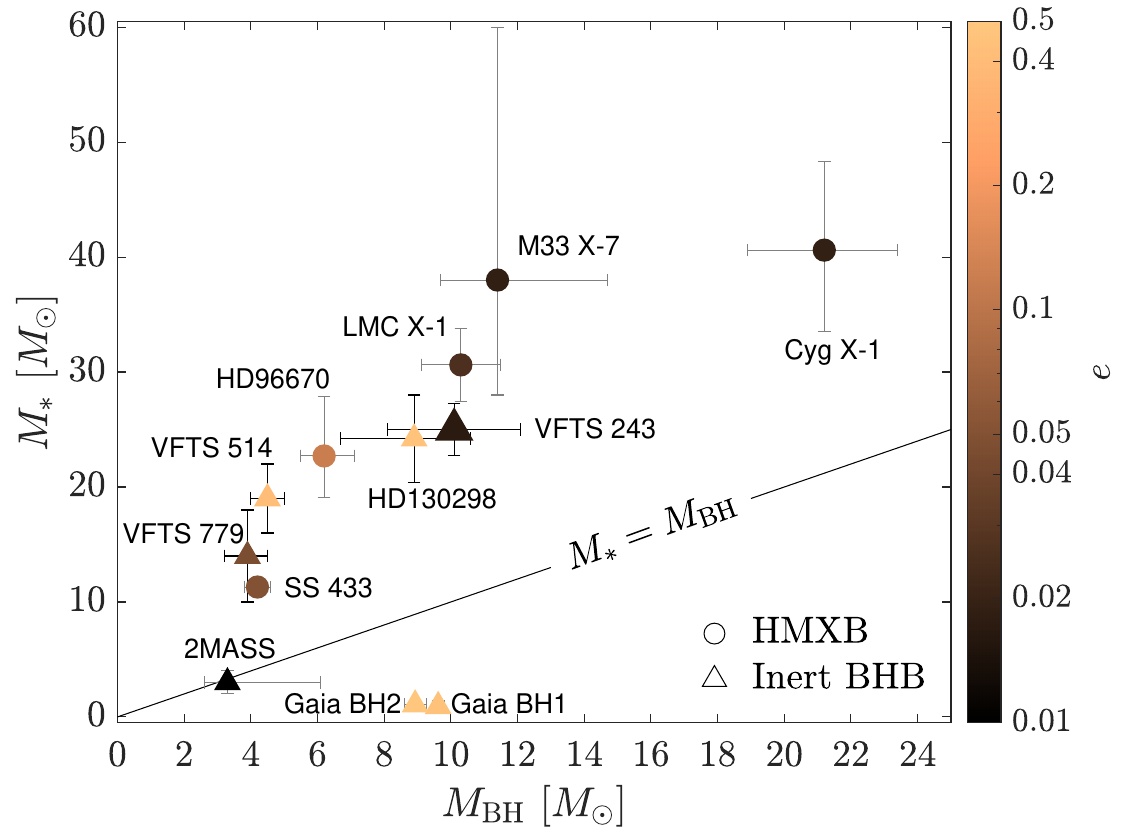}
    \caption{
    Observed sample of field black-hole (BH) binaries with massive companions. For each binary, we show the BH mass ($M_{\rm{BH}}$) on the abscissa, the stellar companion mass ($M_{\rm{*}}$) on the ordinate, and the eccentricity ($e$) through the colorbar. We use circles to indicate high-mass X-ray binaries (HMXBs) and triangles for inert BH binaries (Inert BHB). The symbol of VFTS 243 is slightly larger for clarity.
    }
    \label{fig:black-hole-binaries}
\end{figure}

{\em Properties of VFTS 243.}---VFTS 243 belongs to the family of inert BH binaries recently discovered in the Milky Way \cite{2022A&A...664A.159M} and the Large Magellanic Cloud \cite{2022NatAs...6.1085S} via spectroscopy (Fig.~\ref{fig:black-hole-binaries}). It is comprised of a main-sequence O star with inferred mass of $M_{*}=25.0\pm2.3\ M_{\odot}$ and a  BH companion with $M_{\rm{BH}}=10.1\pm 2.0\ M_{\odot}$, orbital period of $P=10.4031\pm0.0004$~d, and eccentricity $e=0.017\pm 0.012$, where the errors are the 1$\sigma$ uncertainty intervals \cite{2022NatAs...6.1085S}. The relatively high mass of the BH and the nearly circular orbit suggest that the system experienced complete collapse. With a stellar radius ($R$) well within the binary Roche lobe ($f_{\rm{RL}}=R/R_{\rm{RL}}\approx0.33$) \cite{2022NatAs...6.1085S}, VFTS 243 is a relatively wide BH binary (for example, in contrast, Cygnus-X1 is close to filling its Roche lobe with $f_{\rm{RL}}\gtrsim 0.99$ \cite{2021Sci...371.1046M,2021ApJ...908..118N}). Moreover, the super-synchronously rotating massive star of VFTS 243 suggests that the effects of tides in synchronizing and circularizing the orbit are negligible \cite{2022NatAs...6.1085S}.

{\em Constraints on natal kicks.}---In order to analyze the formation of VFTS 243, we use a semi-analytic approach \cite{2002ApJ...574..364P} to calculate the probability that a circular pre-collapse binary that received a natal kick during BH formation could reproduce the orbital configuration of the system as observed today. Models of the pre-collapse binary suggest it initially had a short ($\sim$1 d) orbital period and experienced mass transfer \cite{2022NatAs...6.1085S}. For such configurations, the tidal circularization timescale ($\sim 10^4$ yr \cite{2002MNRAS.329..897H}) is significantly shorter than the time the stars spend on the main sequence ($\sim 10^6$ yr \cite{2018Sci...361.6506F}) prior to BH formation. Therefore, we assume the binary was circularized prior to collapse. Given that the current configuration is barely eccentric, it is reasonable to assume that the orbit was circular and the marginal eccentricity was induced during BH formation. Alternatively, the current eccentricity could be a small residual from incomplete circularization via tides prior to collapse. For the pre-collapse orbital configurations we assume independent, uniform wide prior distributions on the orbital period between $\minPeriod \leq P_i/\text{d} \leq \maxPeriod$, on the natal-kick magnitudes between $\minKick \leq v_{\text{kick}}/(\text{km/s}) \leq  \maxKick$ and on the amount of mass ejected between $\minMass \leq dM/M_\odot \leq \maxMass$. We also assume that the direction of the imparted natal kick is isotropic. We use the post-collapse orbital period, eccentricity and systemic radial velocity as independent constraints (see Supplemental Material), which we justify by the lack of correlation in the observationally inferred orbital parameters \cite{2022NatAs...6.1085S}.

Figure~\ref{fig:results} shows the results of our analysis. The heatmap presents two distinct hotspots. In the limit where $v_{\rm{kick}}\rightarrow0$, we have the solution for pure mass loss (Blaauw effect \cite{1961BAN....15..265B}), where the amount of ejected mass is linearly proportional to the newly acquired eccentricity of the binary. The other spot has $dM\rightarrow0$, i.e., a very small amount of mass is ejected asymmetrically in the frame of reference of the exploding star, and a natal kick makes the orbit eccentric. The marginalized one-dimensional probability distributions have modes at $dM\approx \peakMass\ M_{\odot}$ and $v_{\rm{kick}}\approx \peakVelocity$ km/s. Explosions with moderate mass loss ($\gtrsim1\ M_{\odot}$) and natal kicks ($\gtrsim10$ km/s) can occur (see yellow in the heatmap from Fig.~\ref{fig:results}), but require natal kick directions and magnitudes fine-tuned to balance the mass loss in order to form a low-eccentricity BH binary \cite{2023ApJ...959..106B}, as demonstrated by the diagonal structure of Fig.~\ref{fig:results}. Under the assumption that the binary was circular prior to BH formation and has a small but non-zero eccentricity at present, there is no support for a solution with $dM = v_{\rm{kick}} = 0$.

In Fig.~\ref{fig:results}, the vertical blue band shows the estimated mass decrements associated with the total radiated neutrino energies of collapsing stripped stars \cite{2021ApJ...909..169K}. Numerical simulations of the collapse of massive stars without concomitant supernova explosion in 2D \cite{2022MNRAS.512.4503R} and 3D \cite{2024arXiv240113817J} by the Garching group yield final neutrino-induced BH kick velocities of magnitudes similar to those favored by our analysis of VFTS 243 ($0$--$4$~km/s), in particular also for a pre-collapse stellar model with similar mass. Contesting 3D simulations by the Princeton group found that the final neutrino-induced natal kick velocities of BHs formed via failed supernovae could be larger ($7$--$8$~km/s~ \cite{2023arXiv231112109B}), but still within our 1-$\sigma$ credible intervals. These quantitative differences may be connected to different life times of the NS prior to BH formation or to intrinsic differences of the neutrino asymmetries due to transport.

Under the hypothesis of complete collapse, we perform a back-of-the-envelope estimate on long-term ($\gg1$\,s) neutrino asymmetries during BH formation. We follow Refs. \cite{1994A&A...290..496J,2013MNRAS.434.1355J} and parameterize the linear momentum transferred to the BH as a result of anisotropic neutrino emission as
\begin{equation}
    M_{\rm{BH}}\,v_{\rm{\nu,kick}} = \alpha_{\nu} E^{\rm{tot}}_{\nu}/c = \alpha_{\nu} dM_\nu c\,,
    \label{eq:asymmetry}
\end{equation}
where $\alpha_{\nu}$ is the neutrino emission asymmetry factor, $v_{\rm{\nu,kick}}$ is the neutrino natal kick, $dM_\nu$ is the mass decrement from neutrino emission and $c$ is the speed of light in vacuum. To estimate $\alpha_{\nu}$ and $v_{\rm{\nu,kick}}$ we assume the following (see Supplemental Material for more details). We choose $M_{\rm{BH}}=10\ M_{\odot}$, similar to VFTS 243. We obtain the interval $E^{\rm{tot}}_{\nu}\approx 1$--$11\times 10^{53}\ \rm{erg}$ from the stripped-star BH progenitors from Ref.~\cite{2021ApJ...909..169K}. Finally, the value of $v_{\rm{\nu,kick}}$ is determined by obtaining the subset of the complete natal kick probability distribution that is consistent with complete collapse. These assumptions result in an upper limit on the neutrino asymmetry of $\approx \maxAsymmetryPerCent\%$, where $0\%$ implies spherically symmetric neutrino emission (in the rest frame of the NS), and $33\%$ implies that the dipole component of the neutrino luminosity equals the monopole amplitude \cite{2024arXiv240113817J}. However, the asymmetries are between $\sim 0$--$\likelyAsymmetryPerCent\%$ for the most likely solution of $v_{\nu,\rm{kick}}\approx\peakVelocityNeutrino$ km/s (see Supplemental Material), though a solution with no natal kick is allowed. These values are in overall agreement with the estimates of the anisotropy parameter of the total neutrino emission in recent 3D core-collapse simulations with detailed neutrino transport, namely $0.45$--$0.76\%$ for successful supernova explosions \cite{2024arXiv240113817J} and $0.05$--$0.15\%$ in simulations of non-exploding models \cite{2024arXiv240113817J} (see, however, Ref.~\cite{2022MNRAS.517.3938C}, which reports considerably higher values, probably because of differences in the analysis).

{\em Discussion.}---Our results provide evidence of neutrino-induced natal kicks and add strong support to the complete collapse formation scenario \cite{2022NatAs...6.1085S}. In general, the total natal kick of a compact object has a baryonic, neutrino and gravitational-wave component. If matter ejection in a successful explosion takes place, it generally dominates the total kick for higher-mass progenitors \cite{2020MNRAS.496.2039S,2021ApJ...915...28B,2024arXiv240113817J,2020ApJ...901..108V}, and gravitational waves contribute to the natal kick negligibly, since they carry orders of magnitude less energy with respect to baryons and neutrinos (e.g., \cite{2022MNRAS.512.4503R,2020ApJ...901..108V}). Our calculations provide constraints on the total natal kick and mass loss, which we find to be largely in agreement with mass loss exclusively through neutrino emission and an associated natal kick, rather than baryonic mass ejecta. Depending on the magnitude and direction, a natal kick from baryonic ejecta \cite{2023MNRAS.520.4740S} could further reduce our limit on neutrino natal kicks. Alternatively, there could be a yet-to-be-determined physical mechanism that results in baryonic and neutrino kicks of a similar magnitude that are anti-aligned and almost completely cancel.

Astrometric microlensing has been suggested as a method to constrain BH natal kicks \cite{2020A&A...636A..20W, 2022ApJ...930..159A}; as long as the observed velocity dispersion is much larger than the presumed neutrino kicks, the latter are unconstrained. The entire BH kick distribution cannot be completely explored with bound binaries \cite{2019A&A...624A..66R}, but systems similar to or more massive than VFTS 243 are ideal to explore and constrain low natal kicks. The broad implications of the natal kick that VFTS 243 received during BH formation has been recently investigated in the literature \cite{2022NatAs...6.1085S}, showing support for small \cite{2023MNRAS.520.4740S} and moderate natal kicks \cite{2023ApJ...959..106B}. The solution of very low natal kicks for VFTS 243 is in agreement with neutrino kick estimations from hydrodynamical simulations~\cite{2020PhRvD.101l3013W,2022MNRAS.512.4503R,Kresse2023}. In particular, recent hydrodynamical simulations of core-collapse events yield a range of model-dependent neutrino kicks between $\approx$0--4 km/s in 2D \cite{2022MNRAS.512.4503R} and between $\approx$0--3\,km/s in 3D \cite{Kresse2023,2024arXiv240113817J}. However, other simulations of BH-forming collapses produce neutrino natal kicks between $30\ \rm{km/s} \lesssim v_{\nu,\rm{kick}} \lesssim 100\ \rm{km/s}$ \cite{2022MNRAS.517.3938C}, which would be partially compatible with the tail of our distribution in Fig.~\ref{fig:results} but may be difficult to reconcile with other existing theoretical constraints, if confirmed.

For BH progenitors that have been stripped during a mass transfer episode, the complete collapse scenario suggests almost no baryonic ejecta and a mass decrement exclusively from neutrinos. A very massive proto-NS can lead to a total neutrino energy emission of  $E^{\rm{tot}}_{\nu}\approx 10^{54}$ erg \cite{2021ApJ...909..169K}, which roughly corresponds to a neutrino mass decrement of $dM^{\rm{max}}_{\nu}\approx 0.6\ M_{\odot}$. Therefore $dM < 0.6\ M_{\odot}$, and more realistically  $0.1 \lesssim dM/M_{\odot} \lesssim 0.4$ (see Supplemental Material), is consistent with natal kicks induced  from neutrino emission exclusively \cite{2021ApJ...909..169K}. Alternatively, $dM\gtrsim 0.6\ M_{\odot}$ implies at least some baryonic-mass ejecta.

In addition to VFTS 243, several other inert BH binaries have been recently reported: HD 130298 \cite{2022A&A...664A.159M}, Gaia BH1 \cite{2023MNRAS.518.1057E,2023AJ....166....6C} and Gaia BH2 \cite{2023MNRAS.521.4323E}, as well as inert BH binary candidates 2MASS J05215658+4359220 \cite{2019Sci...366..637T,2020Sci...368.3282V}, HD 96670 \cite{2021ApJ...913...48G}, VFTS 514 \cite{2022A&A...665A.148S}, VFTS 779 \cite{2022A&A...665A.148S}, all of them presented in Fig.~\ref{fig:black-hole-binaries}.  Arguably, the most atypical systems in the sample are the astrometric BH binaries Gaia BH1 and Gaia BH2.  With massive BHs ($M_{\rm{BH}}\approx10\ M_{\odot}$), they are at wide orbital periods ($\gtrsim 100$--$1000$ d), have low-mass ($M_{*}\approx1\ M_{\odot}$) companions and possibly formed dynamically \cite{2024MNRAS.527.4031T,2023arXiv230613121D}, which could explain their large eccentricity ($e\approx 0.5$), putting them in a different class with respect to the other BH binaries, whose assembly is more in line with that of stellar binaries \cite{2020A&A...638A..39L}.

The sample of single line spectroscopic (SB1) inert BH binaries is diverse (Fig.~\ref{fig:black-hole-binaries}). VFTS 243 is the most massive and least eccentric of SB1 BH binaries. HD 130298 is slightly less massive than VFTS 243 (with $M_{\rm{BH}}\approx 9\ M_{\odot}$ and $M_{*}\approx24\ M_{\odot}$), has a similarly small Roche-lobe filling factor ($f_{\rm{RL}}=0.26$) and a similar orbital period ($P_{\rm{orb}}\approx 15$ d), but  it is quite eccentric ($e\approx0.5$). VFTS 514 is even less massive (with $M_{\rm{BH}}\approx5\ M_{\odot}$ and $M_{*}\approx19\ M_{\odot}$) and is also quite eccentric ($e\approx0.4$); with a long orbital period ($P_{\rm{orb}}\approx185$ d), it is likely far from Roche-lobe overflow. VFTS 779 is the least massive of the SB1 BH binaries (with $M_{\rm{BH}}\approx4\ M_{\odot}$ and $M_{*}\approx14\ M_{\odot}$), has a near circular  orbit ($e\approx 0.02$) and a long orbital period ($\approx 60$ d). The spectral type of the star of VFTS 779 indicates it has evolved past the main sequence \cite{2022A&A...665A.148S}; at this stage, it is developing a convective envelope which makes tidal circularization orders of magnitude more efficient than for main-sequence stars with radiative envelopes (see also discussion in Supplemental Material).

The sample of BH HMXBs \cite{2023A&A...671A.149F,2023A&A...677A.134N} is broader in the mass parameter space, but otherwise is more homogeneous (Fig.~\ref{fig:black-hole-binaries}). Most BH HMXBs have short orbital periods ($P_{\rm{orb}}<6$ d), small eccentricities ($e\lesssim 0.1$) and large Roche-filling factors ($f_{\rm{RL}}\gtrsim0.9$). Large Roche-filling factors in HXMBs are in agreement with the theory of wind-accreting X-ray binaries, which predicts a threshold of $f_{\rm{RL}}\gtrsim0.8$--$0.9$ as a condition for observability \cite{2021PASA...38...56H}. Cygnus X-1 is the archetype of BH HMXBs, with $M_{\rm{BH}}\approx 21\ M_{\odot}$ and $M_{*}\approx 41\ M_{\odot}$, a short-day orbital period ($P_{\rm{orb}}\approx 6$ d), a large Roche-lobe filling factor ($f_{\rm{RL}}>0.99$), and a near-circular orbit ($e\approx0.02$). While the eccentricities of Cygnus X-1, LMC X-1 and M33 X-1 are similar to that of VFTS 243, both wind accretion onto the compact object \cite{2002MNRAS.329..897H,2012ApJ...747..111W} and tides could have played a role in circularizing these BH HMXBs. For massive, main-sequence stars with radiative envelopes, the dynamical tide is the dominant tidal dissipation mechanism \cite{1975A&A....41..329Z}. Dynamical tides are not an efficient dissipation mechanism for VFTS 243 \cite{2022A&A...665A.148S}, the most compact inert BH binary detected to date. The circularization timescale via dynamical tides is a strong function of the separation ($\tau_{\rm{circ}}\propto (a/R)^{21/2}$ \cite{1977A&A....57..383Z}, where $a$ is the semi-major axis of the binary), which implies that tides are significantly less efficient for the SB1 sample than for the observed BH HMXBs. For example, doubling the orbital period of Cygnus X-1 would result in a similar period to VFTS 243 and would reduce the impact of tides by $\sim2$ orders of magnitude. Moreover, HD 130298 has rather similar properties (masses, periods and Roche-filling factors) to VFTS 243, but is rather eccentric; this demonstrates that dynamical tides are not efficient at circularizing the orbit at such large separations.

Finally, we highlight that for BH binaries with $M_{*}>M_{\rm{BH}}$ there seems to be  a sharp drop in eccentricity for systems with $M_{\rm{BH}}\gtrsim 10\ M_{\odot}$. We speculate that this drop could denote the transition between luminous, mass-shedding explosions and complete collapse \cite{2001ApJ...554..548F,2013ApJ...769..109L,2015MNRAS.450.3289G,2017MNRAS.468.4968A}. The early theory of stellar collapse suggested that less massive cores eject some mass during BH formation, while complete collapse could only occur for cores of mass $\gtrsim 11\ M_{\odot}$ \cite{2012ApJ...749...91F}. However, a growing body of theoretical work shows evidence that the compact object remnant mass function is a non-monotonic function of mass \cite{2011ApJ...730...70O,2012ApJ...757...69U,2012ARNPS..62..407J,2015ApJ...801...90P,2016ApJ...821...38S,2016ApJ...818..124E,2016ApJ...821...69E,2016MNRAS.460..742M,2019ApJ...870....1E,2020ApJ...890..127C,2021Natur.589...29B,2023MNRAS.526.5900V} that could even be stochastic \cite{2020MNRAS.499.3214M}, dependent on initial rotation and composition \cite{2018ApJS..237...13L} and that might be affected by binary interactions \cite{2021A&A...645A...5S,2021ApJ...916L...5V,2021A&A...656A..58L}. This means that more massive stars do not always result in more massive compact objects; some stellar cores above a certain threshold can result in mass-shedding explosions and form NSs \cite{2020ApJ...890...51E}. Based on the sharp eccentricity transition between HD 130298 and VFTS 243 (both SB1 BH binaries with similar masses and orbital periods) and the results presented in this \textit{Letter}, we consider that helium cores with a mass of $\approx 10\ M_{\odot}$ could undergo complete collapse. However, this does not rule out the possibility that complete collapse occurs for less massive stars, or that incomplete collapse returns at higher masses. Future observations should provide more precise values on the actual mass threshold for complete collapse.

{\em Conclusions.}---We explore the effect of mass loss and natal kicks during BH formation in VFTS 243, the heaviest inert stellar-mass BH binary detected to date. Our results show that the marginalized distributions peak around $dM=\peakMass\ M_{\odot}$ and $v_{\rm{kick}}=\peakVelocity$ km/s, respectively; these values are consistent with recent simulations of the stellar collapse predicting mass loss via neutrino emission exclusively  \cite{2021ApJ...909..169K}, fully in agreement with the complete collapse scenario \cite{2011ApJ...730...70O}. Under the hypothesis of complete collapse, we estimate the asymmetry in the neutrino emission to be $\sim0$--$\likelyAsymmetryPerCent\%$ and provide an upper limit of $\approx \maxAsymmetryPerCent\%$. We find that mass loss $\lesssim \massOneSigma\ M_{\odot}$ and $v_{\rm{kick}}\lesssim \kicksOneSigma\ \rm{km/s}$ are preferred at a $68\%$ C.L., in agreement with other studies in the literature \cite{2022NatAs...6.1085S,2023MNRAS.520.4740S}. Moreover, our results also accommodate solutions with no natal kick and no asymmetries in neutrino emission. The progenitor of the BH component of VFTS 243 is likely a massive stripped star \cite{2022A&A...665A.148S} which experiences negligible mass loss during BH formation. Following recent observations of inert BH binaries in the Local Group (Fig.~\ref{fig:black-hole-binaries}), we suggest that complete collapse can occur for stars that end their lives with cores of $\gtrsim 10\ M_{\odot}$.

{\em Acknowledgments.}---We thank David Aguilera-Dena, Sharan Banagiri, Sebastian Gomez, Laurent Mahy, Thomas Maccarone, Bernhard Mueller, Rachel Patton and Ruggero Valli for useful discussions. In Garching, this work was supported by the German Research Foundation (DFG) through the Collaborative Research Centre ``Neutrinos and Dark Matter in Astro- and Particle Physics (NDM),'' Grant SFB-1258\,--\,283604770, and under Germany's Excellence Strategy through the Cluster of Excellence ORIGINS EXC-2094-390783311. RW acknowledges support from the KU Leuven Research Council through grant iBOF/21/084. IT acknowledges support from the Villum Foundation (Project No.~37358), the Carlsberg Foundation (CF18-0183), and the Danmarks Frie Forskningsfonds (Project No.~8049-00038B).

\bibliography{neutrinos}

\onecolumngrid
\appendix

\newpage
\clearpage
\setcounter{equation}{0}
\setcounter{figure}{0}
\setcounter{table}{0}
\setcounter{page}{1}
\makeatletter
\renewcommand{\theequation}{S\arabic{equation}}
\renewcommand{\thefigure}{S\arabic{figure}}
\renewcommand{\thepage}{S\arabic{page}}

\begin{center}
\textbf{\large Supplemental Material\\[0.5ex]
Observational evidence for neutrino natal kicks from black-hole binary VFTS 243
}
\end{center}
In this Supplemental Material, we first describe the modeling of the natal kicks, discuss the caveats in our model, and compare with other studies in the literature. Then, we describe our analysis on the estimates of the neutrino asymmetries in the context of the black hole (BH) of VFTS 243. Finally, we present a more detailed description on the binary evolution leading to the formation and fate of VFTS 243, as well as justify the assumptions made in our analysis.

\section{Modeling of natal kicks}
Evidence for large neutron star (NS) natal kicks comes from observations of young, isolated pulsars with speeds of the order of a few $100$ km/s, which are assumed to reflect their speeds at birth \cite{2005MNRAS.360..974H,2017A&A...608A..57V,2021ApJ...920L..37W}.  If the natal kick is mainly due to asymmetric baryonic mass loss, any fallback leading to BH formation is expected to suppress the natal kick. For modeling purposes, the natal-kick reduction is often assumed for simplicity to scale proportionally with the fraction of the envelope material that falls back onto the nascent BH \cite{2012ApJ...749...91F,2013MNRAS.434.1355J}. Ref.~\cite{2019MNRAS.489.3116A} finds that BHs in X--ray binary systems could have natal kicks up to a few $100$ km/s, following a possibly bimodal distribution.

We modeled the effects of a supernova in a binary system with arbitrary natal kick and mass loss based on Ref~\cite{2002ApJ...574..364P}. We do not distinguish  baryonic ($dM_{\rm{bar}}$) from neutrino ($dM_{\nu}$) mass losses, so our resulting upper limits apply to $dM = dM_{\rm{bar}} + dM_{\nu}$. Similarly, upper limits are placed on the overall natal kick $v_{\text{kick}}$, which is the magnitude of the vector sum of any hydrodynamic and neutrino natal kicks. The momentum radiated in gravitational waves is negligible \cite{2019MNRAS.490.5560H} and we therefore ignore it in the natal-kick analysis. We assume the orbit was circular prior to the supernova, and that there is no change to the mass or instantaneous velocity of the companion star.

\begin{figure}
    \centering
    \includegraphics[trim=13cm 0 0 0, clip,width=\columnwidth]{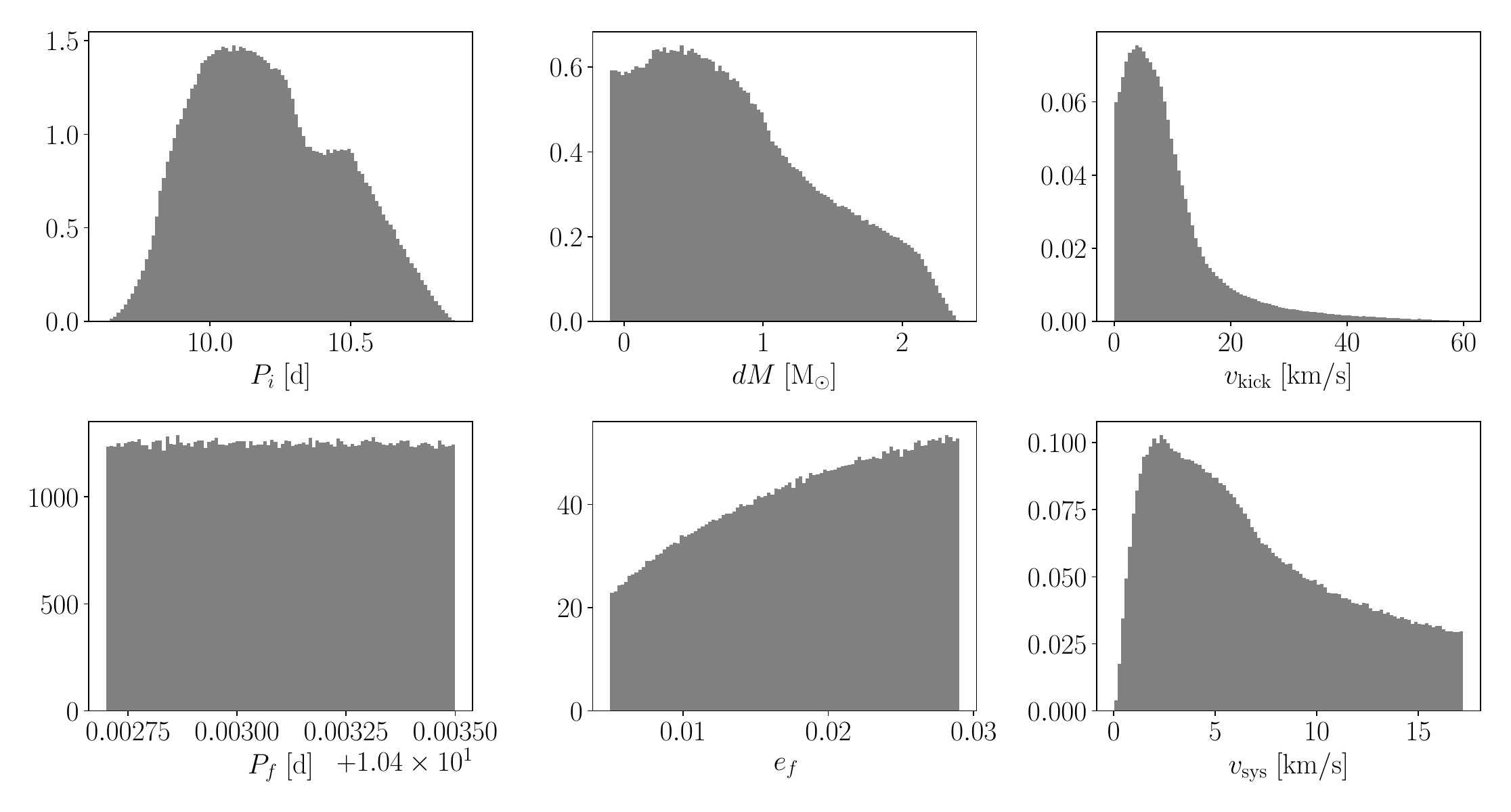}
    \caption{
    Normalized distributions of the properties of binaries resulting in VFTS 243. We show the pre-collapse orbital period ($P_i$), mass decrement after collapse ($dM$), and natal kick magnitude ($v_{\rm{kick}}$) in the top row. The post-collapse orbital period ($P_f$), post-collapse eccentricity ($e_f$) and systemic velocity ($v_{\rm{sys}}$) plotted in the bottom row are constrained to match the reported values of VFTS 243.
    }
    \label{fig:statistics}
\end{figure}

Fixing the present-day masses for a given binary, we sample from a set of independent initial distributions on $dM$, $v_{\text{kick}}$, and the pre-collapse orbital period $P_i$. In our analysis, we assume independent uniform initial distributions on all three parameters, with ranges between $\minMass \leq dM/M_\odot \leq \maxMass$, $\minKick \leq v_{\text{kick}}/(\text{km/s}) \leq \maxKick$, and $\minPeriod \leq P_i/\text{d} \leq \maxPeriod$. The final joint distribution is marginalized over $P_i$. The kick direction is taken to be isotropic in the frame of the collapsing star. We construct a probability for each triplet $(dM, v_{\text{kick}}, P_i)$ from the fraction of the sampled kick directions which result in a final orbital period $P_f=10.4031\pm0.0004$ d and eccentricity $e_f=0.017\pm0.012$, where the errors are the 1$\sigma$ uncertainty intervals provided in the detection paper of VFTS 243 \cite{2022A&A...665A.148S}. We use the systemic radial velocity as an additional constraint. The systemic radial velocity of VFTS 243 is $260.2\pm0.9$ km/s, as reported by the discovery paper \cite{2022A&A...665A.148S}. The 1$\sigma$ of the measured mean of the Tarantula region of $271.6\pm12.2$ km/s suggests a low ($\sim 1$--$10$ km/s) systemic radial velocity. We only consider systems with systemic velocity of less than $\CoMKicksOneSigma$ km/s, which correspond to the 68\% percentile on the absolute value of the difference of the two normal distributions for VFTS 243 and the Tarantula region. From our analysis, the natal kicks that result in a VFTS 243-like binary have a systemic velocity distribution that peaks around  $\peakSystemicVelocity$ km/s (Fig.~\ref{fig:statistics}).

In Fig.~\ref{fig:statistics}, we show the initial and final distributions leading to $10^6$ VFTS 243-like systems. We find that the choice of the range of the initial distributions completely span the pre-supernova parameter-space of interest leading to the configuration of VFTS 243 as observed today, i.e., our analysis is not constrained by the chosen prior boundaries. The impact on the period and eccentricity from significant mass loss can be offset by a large and fortuitously directed natal kick; this fortuitousness is quantified by the solid angle on the kick direction unit sphere within which the final orbital properties are consistent with observations, which is precisely the probability we estimate. For very large $dM$ and $v_{\text{kick}}$, the probability of producing a low eccentricity orbit is vanishingly small, providing a fairly tight constraint on both $dM$ and $v_{\text{kick}}$, as shown in Fig.~\ref{fig:results}.

In the context of near-circular BH binaries, the scenario of an eccentric pre-collapse binary is unlikely, but cannot be ruled out. From observations of short-period ($<10$ d) massive O-type binaries on the main sequence, all of them have $e<0.4$, and approximately half have $e<0.1$ (cf.~Fig.~3 of Ref.~\cite{2017ApJS..230...15M}). Moreover, VFTS 243 is likely to have experienced a mass transfer episode where the binary could have been further circularized \cite{2022A&A...665A.148S}, either via tidal interactions, gas drag or both. In the context of Cygnus X-1, the plausibility of an eccentric pre-collapse orbit was explored and concluded to be unlikely \cite{2012ApJ...747..111W}; we expect something similar for VFTS 243. Moreover, an eccentric pre-collapse orbit leading to a circular post-collapse orbit requires a fortuitous kick. This rare fortuitous kick would likely be non-negligible in magnitude, broadening the natal kick distributions presented in Fig.~\ref{fig:results}. We consider the scenario of a near circular pre-supernova orbit as the more likely one, given that it requires less fine tuning. On the other hand, it is possible that some of the observed eccentricity is due to a small residual from the pre-collapse orbit rather than a neutrino kick, which would further reduce our neutrino mass loss and kick estimates.

We now briefly contrast the impact of our assumptions on the natal-kick magnitude distribution with the ones in previous work on VFTS 243 \cite{2022NatAs...6.1085S,2023MNRAS.520.4740S,2023ApJ...959..106B}. Ref.~\cite{2023MNRAS.520.4740S} explores the impact of a uniform prior---consistent with our approach---with that of a Maxwellian prior for the natal kick magnitude. They find that the Maxwellian distribution leads to a slightly larger posterior in the natal-kick parameter space. This is likely because the Maxwellian distribution rarely allows for very low natal kicks, which are the preferred solution in the VFTS 243 configuration. Moreover, the Maxwellian natal-kick distribution is motivated by observations of isolated pulsars \cite{2005MNRAS.360..974H} and therefore might not be adequate for black holes formed via fallback or complete collapse; in fact, the apparent Maxwellian distribution of NS velocities may be an accidental consequence of the integration of a physical distribution of individual kicks over all progenitor systems \cite{2023MNRAS.519.5893K}. We consider the uniform natal-kick magnitude prior to be an agnostic and adequate choice. Ref.~\cite{2023ApJ...959..106B} perform a full Bayesian analysis, using wide priors for all parameters and comparing uniform and Maxwellian velocity priors. Their results are in agreement with our work and the results from Ref.~\cite{2023MNRAS.520.4740S}, once the use of different (90\%) confidence intervals and large ($>5\ M_{\odot}$) mass ejecta is accounted for. Finally, they show that the eccentricity induced by large ejecta can be canceled by a strong, fortuitously aligned natal kick; however, this outcome is rather unlikely.

\section{Estimating neutrino asymmetries}
We have explored the natal-kick parameter space of VFTS 243 and found that low-mass loss ($dM\approx\peakMass\ M_{\odot}$) and low natal-kick magnitudes ($v_{\rm{kick}}\approx\peakVelocity$ km/s) are the preferred parameters to match VFTS 243 as observed today (Fig.~\ref{fig:results}). However, there are also less likely, but plausible, $dM$--$v_{\rm{kick}}$ combinations with larger mass loss and natal kicks that could explain the current orbital configuration of VFTS 243. Here we consider exclusively the $dM$--$v_{\rm{kick}}$ pairs that are in agreement with models where there are no baryonic mass ejecta and the mass decrement and natal kick come exclusively from neutrinos, which implies $v_{\nu,\rm{kick}}=v_{\rm{kick}}$. In practice, this means that we select a subset of the complete natal kick distribution that is within the blue shaded region from Fig. 1. Under this assumption, we calculate the asymmetry in the neutrino emission in the BH of VFTS 243.

\begin{figure}
    \centering
    \includegraphics[trim=1cm 7cm 1cm 7cm, clip,width=0.48\columnwidth]{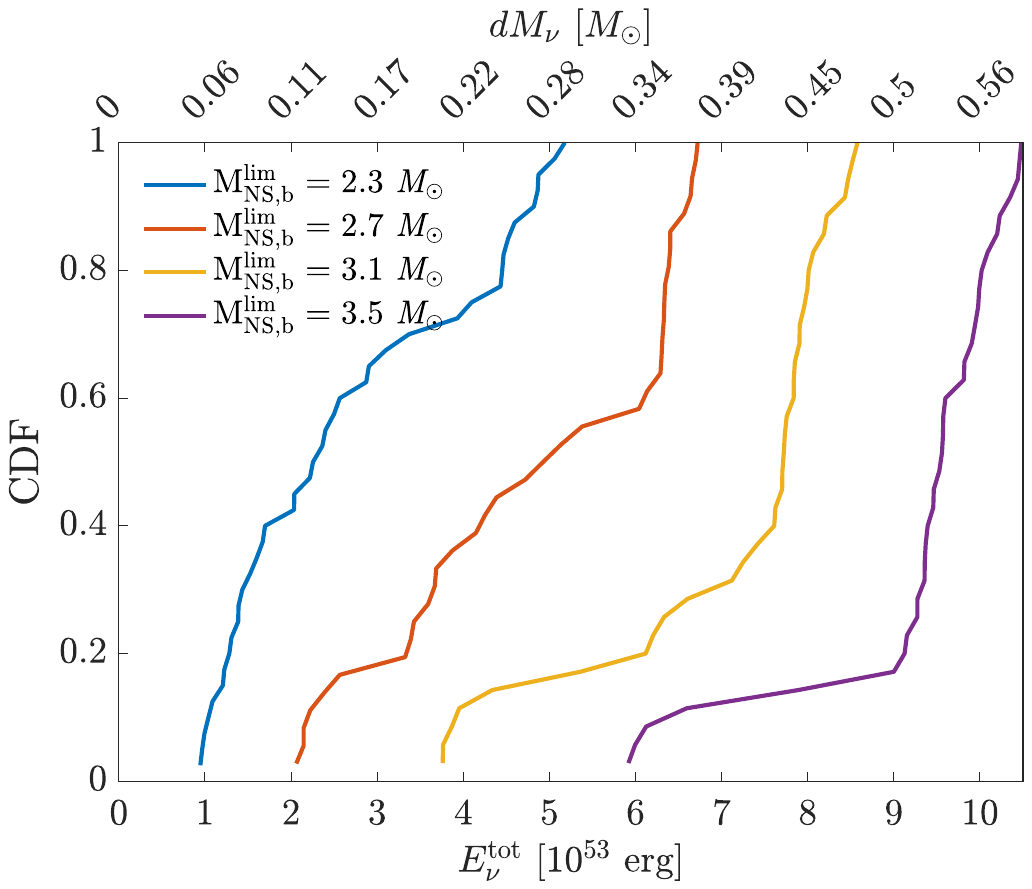}
    \caption{
    Cumulative distribution function (CDF) of the total neutrino energy emission ($E^{\rm{tot}}_{\nu})$ for a selected sample of the stellar models explored in Ref.~\cite{2021ApJ...909..169K}. This sample includes exclusively helium models, a proxy for stripped stars, which result in BH formation. We present the solutions for different NS baryonic mass limits $\rm{M_{NS,b}^{lim}}=\{2.3, 2.7, 3.1,3.5\}\ M_{\odot}$ in blue, red, yellow and purple, respectively.
    }
    \label{fig:neutrino-energies}
\end{figure}

We use the results from Ref.~\cite{2021ApJ...909..169K}, which investigates successful and failed supernova explosions with parameterized one-dimensional neutrino engines \footnote{These results are available upon request in the Garching Core-Collapse Supernova Archive \url{https://wwwmpa.mpa-garching.mpg.de/ccsnarchive/}}. While they explore a large set of stellar models, we only consider the results of their helium-star models; the progenitor of VFTS 243 and similar inert BH binaries probably experienced a mass transfer episode where their BH progenitor was stripped \cite{2020A&A...638A..39L}, and therefore helium-star progenitors that lead to BH formation are most relevant. From these models, we retrieve the total energy radiated in all species of neutrinos, which is bounded between $1\lesssim E^{\rm{tot}}_{\nu}/10^{53} \rm{erg}\lesssim 11$ (Fig.~\ref{fig:neutrino-energies}). This energy is strongly correlated with the NS baryonic mass upper limit $M_{\rm{NS,b}}^{\rm{lim}}$. For our analysis, we use the complete set of values available from Ref.~\cite{2021ApJ...909..169K}, which are $M_{\rm{NS,b}}^{\rm{lim}}=\{2.3,2.7,3.1,3.5\}\ M_{\odot}$ (Fig.~\ref{fig:neutrino-energies}); the largest values of $M_{\rm{NS,b}}^{\rm{lim}}$ are very high, in tension with observations of maximum NS masses \cite{2017ApJ...850L..19M,2020PhRvD.101f3021S}. The range of total radiated neutrino energies is mostly sensitive to $M_{\rm{NS,b}}^{\rm{lim}}$ and not so sensitive to the pre-supernova core mass; for example, further constraining the models to pre-supernova masses similar to VFTS 243 would barely change the range of total neutrino energies.

After we have obtained the values of $E^{\rm{tot}}_{\nu}$, we estimate the mass decrement associated with this emitted energy as $E^{\rm{tot}}_{\nu}/c^2 = dM_{\nu}$. Our estimates for the neutrino-mass decrement yield values between $0.1\lesssim dM_{\nu}/M_{\odot} \lesssim 0.6$, shown as the blue shaded region in Fig.~\ref{fig:results}.

\begin{figure}
    \centering
    \includegraphics[width=0.48\columnwidth]{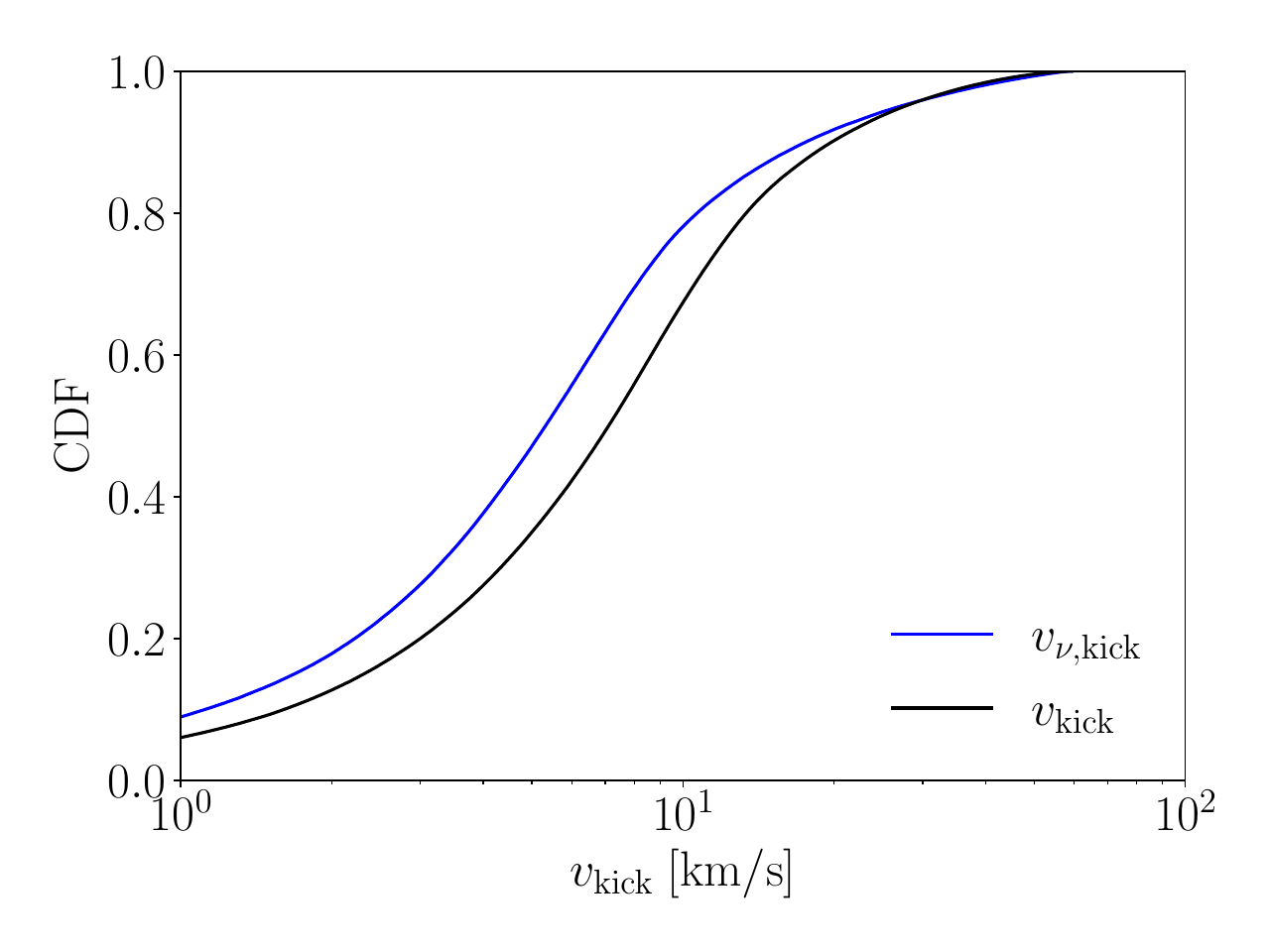}
    \includegraphics[width=0.48\columnwidth]{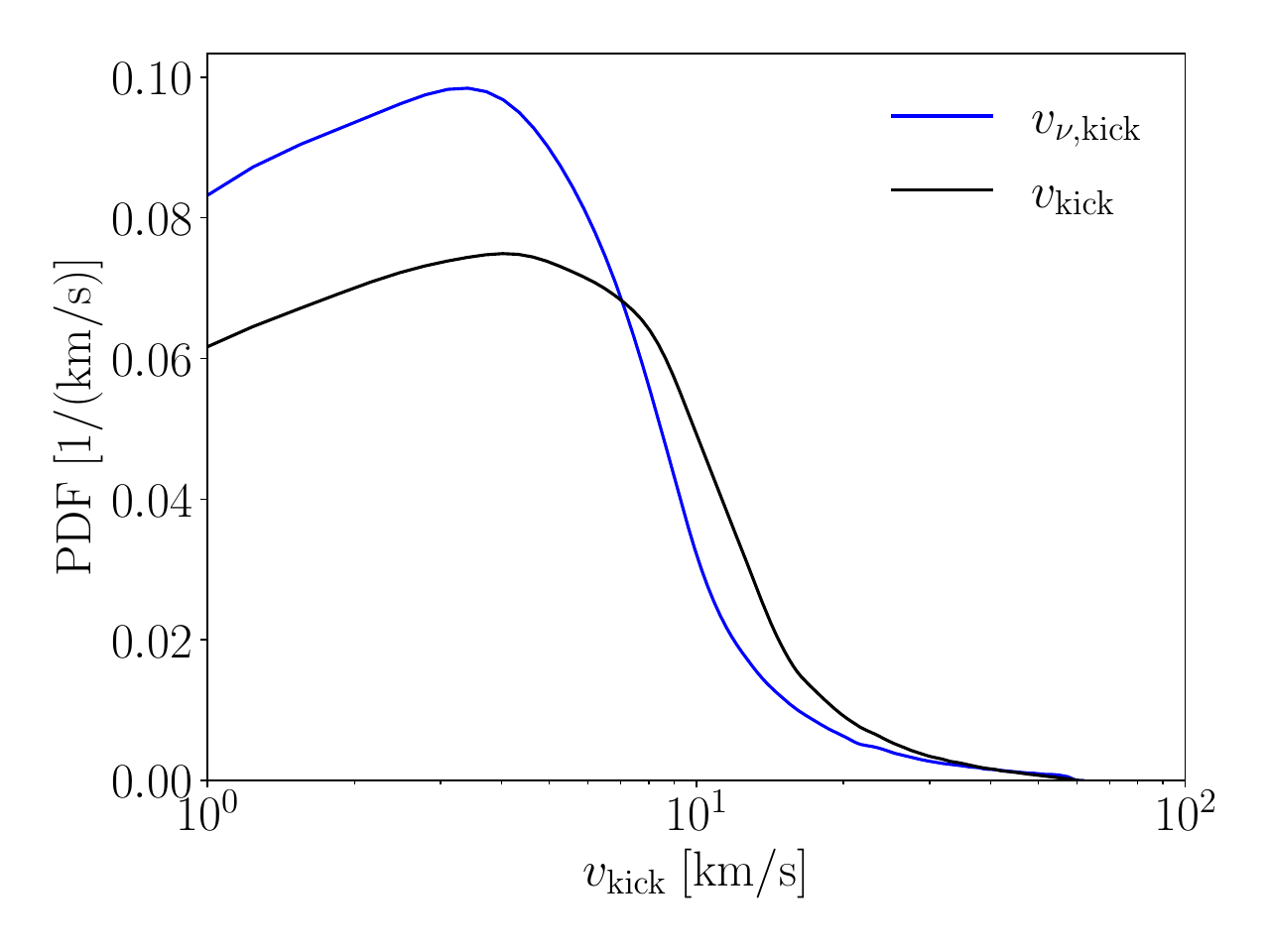}
    \caption{
    Natal-kick distribution for VFTS 243-like binaries. The complete distribution ($v_{\rm{kick}}$), as presented in the main text, is shown as a solid black line. The blue line corresponds to natal-kick velocity distributions where the mass decrement can be associated exclusively with neutrino emission ($v_{\rm{\nu,kick}}$); i.e. they are a subset of the complete distribution for values in mass decrement that are supported by the theoretical models from Ref.~\cite{2021ApJ...909..169K}. The left panel shows cumulative distribution functions (CDF). For each distribution we perform kernel density estimation (KDE) in order to estimate the smooth probability distribution function (PDF); the PDFs are normalized independently and shown in the right panel. The neutrino natal kick PDF peaks at $\peakVelocityNeutrino$ km/s, only slightly below the complete kick distribution PDF, which peaks close to $\peakVelocity$ km/s.
    }
    \label{fig:neutrino-natal-kicks}
\end{figure}

We use our estimates of the neutrino-mass decrement to create a neutrino natal-kick magnitude distribution from the probability estimates presented in Fig.~\ref{fig:results}. In practice, this means that for each value of $M_{\rm{NS,b}}^{\rm{lim}}$ we obtain the limits on $E^{\rm{tot}}_{\nu}$ (Fig.~\ref{fig:neutrino-energies}) and filter the natal-kick distribution between $dM_{\nu}^{\rm{min}} \leq dM \leq dM_{\nu}^{\rm{max}}$, which assumes $dM = dM_{\nu}$. This subset of the natal-kick probability distribution has preference for lower natal kicks with respect to the complete distribution (Fig.~\ref{fig:neutrino-natal-kicks}); this distribution peaks at $\approx \peakVelocityNeutrino$ km/s and has a $\rm{CDF}(\textit{v}_{\rm{\nu,kick}}=\nuKicksOneSigma\ \rm{km/s})\approx 0.68$ (Fig.~\ref{fig:neutrino-natal-kicks}), in contrast with a peak at $\approx \peakVelocity$ km/s and $\rm{CDF}(\textit{v}_{\rm{kick}}=\kicksOneSigma\ \rm{km/s})\approx 0.68$ when using the complete sample.

\begin{figure}
    \centering
    \includegraphics[trim=0cm 7.5cm 0cm 8.0cm, clip,width=0.5\columnwidth]{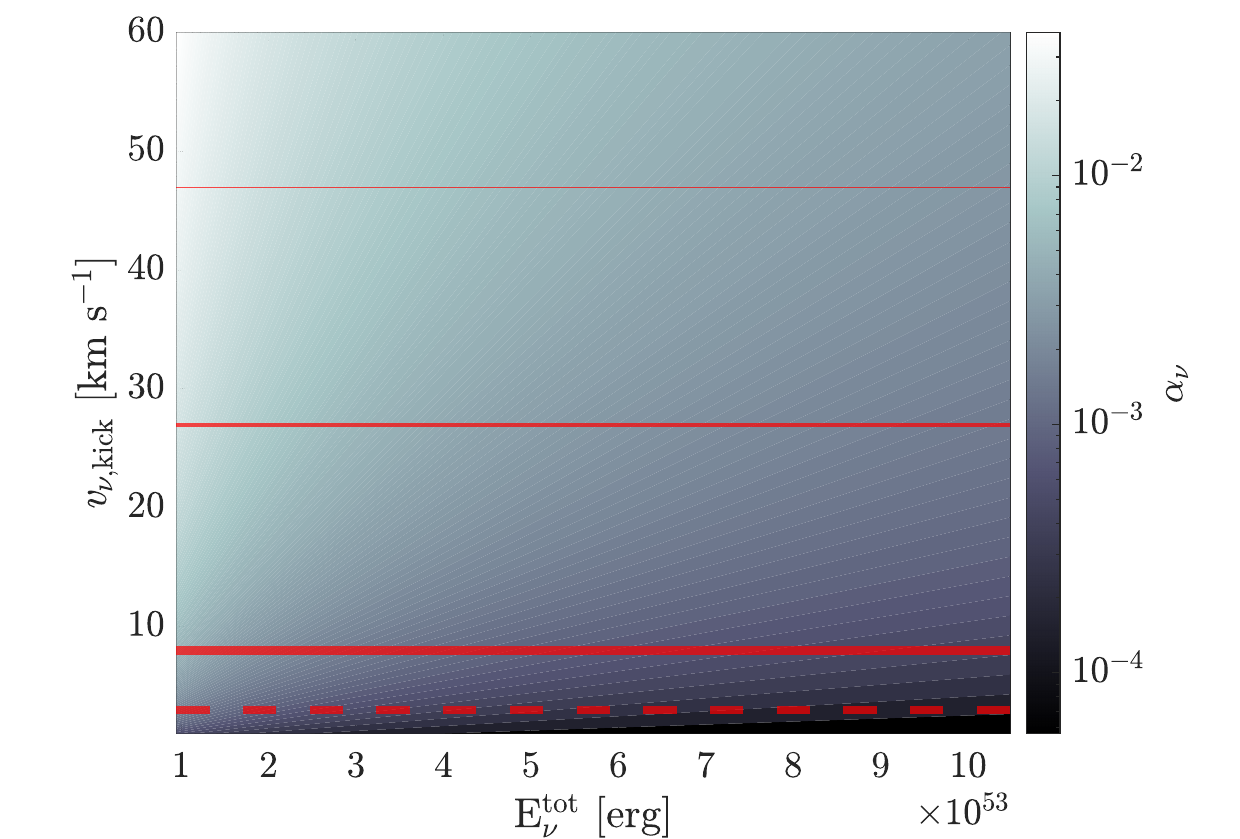}
    \caption{
    Exploration of the neutrino emission asymmetry parameter ($\alpha_{\nu}$) in the neutrino natal-kick magnitude ($v_{\nu,\rm{kick}}$) and total neutrino energy emission ($E_{\nu}^{\rm{tot}}$) parameter space. We follow Eq. 1 to explore the asymmetries for a black hole with mass $10\ M_{\odot}$, similar to VFTS 243 \cite{2022NatAs...6.1085S}. The abscissa is bound by $E_{\nu}^{\rm{tot}}$ as calculated by Ref.~\cite{2021ApJ...909..169K} (see also Fig.~\ref{fig:neutrino-energies}). The ordinate is bound by the maximum value of $v_{\nu,\rm{kick}}$ (see Fig.~\ref{fig:neutrino-natal-kicks}). The horizontal dashed red line shows the peak of the neutrino natal-kick distribution (see Fig.~\ref{fig:neutrino-natal-kicks}). The horizontal solid red lines show the values of $v_{\nu,\rm{kick}}$ where $\rm{CDF}(\textit{v}_{\nu,\rm{kick}})=0.68$ (thick), $\rm{CDF}(\textit{v}_{\nu,\rm{kick}})=0.95$ (regular) and $\rm{CDF}(\textit{v}_{\nu,\rm{kick}})=0.99$ (thin).
    }
    \label{fig:neutrino-asymmetries}
\end{figure}

Finally, we proceed to calculate the neutrino emission asymmetry factor $\alpha_{\nu}$. We follow Refs. \cite{1994A&A...290..496J,2013MNRAS.434.1355J} and parameterize the linear momentum transferred to the BH as a result of anisotropic neutrino emission as
\begin{equation}
    M_{\rm{BH}}\,v_{\rm{\nu,kick}} = \alpha_{\nu} \frac{E^{\rm{tot}}_{\nu}}{c} \,,
\end{equation}
where $c$ is the speed of light in vacuum. We assume $M_{\rm{BH}}=10\ M_{\odot}$ as in VFTS 243 \cite{2022NatAs...6.1085S}. We present the results of this analysis in Fig.~\ref{fig:neutrino-asymmetries}. The largest value of the asymmetry is $\alpha_{\nu}^{\rm{max}}=\maxAsymmetryFactor$, which corresponds to a natal-kick of $\approx \nuKickOneSigmaMax$ km/s (at 68\% C.L.) , limited by the largest natal kick from our model which is able to reproduce VFTS 243 as observed today. The natal kick is limited by the constraints on the systemic radial velocity. For the most likely value of the natal kick distribution, $v_{\rm{\nu,kick}}=\peakVelocityNeutrino$ km/s (Fig.~\ref{fig:neutrino-natal-kicks}), the asymmetries are approximately between $\minAsymmetryFactorLimForMaxLikelihood \lesssim \alpha_{\nu} \lesssim \maxAsymmetryFactorLimForMaxLikelihood$. For natal-kicks of $\lesssim \nuKicksOneSigma$ km/s, where $\rm{CDF}(\textit{v}_{\rm{\nu,kick}}=\nuKicksOneSigma\ \rm{km/s})\approx 0.68$, the asymmetries are $\alpha_{\nu} \lesssim \lastAsymmetryFactorLimForMaxLikelihood$.

\section{Binary evolution leading to VFTS 243}
While the precise history of the formation and evolution of VFTS 243 are uncertain, there is overall consensus on the progenitors and fate of OB star-BH systems \cite{2020A&A...638A..39L}. The default model of Ref.~\cite{2022NatAs...6.1085S} reports the following initial conditions to form VFTS 243. An $\approx 4$ d massive detached circular binary, comprised of non-rotating stars with primary and secondary masses of $\approx 30\ M_{\odot}$ and $\approx 22\ M_{\odot}$, respectively, evolves until the primary fills its Roche lobe and a mass transfer episode begins \cite{2022NatAs...6.1085S}. In this mass transfer episode, the primary is stripped of its hydrogen-rich envelope that can be partially or completely accreted by the secondary. The now stripped primary continues its evolution until it collapses to a BH, which is the configuration we observe today. We proceed to describe some of the several uncertainties in the evolutionary channel just introduced.

Many initial conditions could result in the current configuration of VFTS 243. The eccentricity distribution of close ($\lesssim 10$ d) OB-type binaries suggest that they are never extremely eccentric ($e<0.4$) and that they have moderate eccentricities ($e\approx 0.1-0.4$) or are effectively circular \cite{2012Sci...337..444S,2017ApJS..230...15M}. The initial orbital period determines whether the first mass transfer episode will occur while the primary is on the main sequence (case A) or just after it has left the main sequence (case B). Case A mass transfer is expected for orbital periods $\lesssim 10$ d; alternatively case B mass transfer is expected for longer orbital periods \cite{2020A&A...638A..39L}. This first mass transfer episode can significantly alter the secondary star, which becomes spun-up, rejuvenated and polluted \cite{2021ApJ...923..277R}. Moreover, the fraction of the donated mass that is accreted, i.e. the \textit{conservativeness} of the mass transfer episode, directly impacts the orbital evolution; this conservativeness is one of the most poorly constrained parameters in massive binary evolution, along with the specific angular momentum lost during non-conservative mass transfer \cite{2023ApJ...958..138W}. After the primary has been stripped, the time before it collapses into a BH is relatively short with respect to the duration of the main sequence. If the BH forms via complete collapse, the orbit is not significantly affected, as described in this \textit{Letter}. Now that the primary has formed a BH, the evolution of the secondary determines the evolution of the system. There are three main physical processes that govern the evolution of detached binaries, or in this case inert BH binaries: gravitational waves, stellar winds and tides. Gravitational waves dissipate energy and result in the inspiral and circularization of the binary \cite{1964PhRv..136.1224P}. For systems as wide as VFTS 243, gravitational waves are an extremely inefficient dissipation mechanism and their effect is negligible. Stellar winds can generally widen the binary, but hydrodynamical gas drag can play a role if the wind velocity is comparable to the orbital velocity \cite{2021arXiv210709675S}. In order to determine the role of gas drag we compare the terminal wind speed of $v_{\rm{wind}}\gtrsim 2000\ \rm{km/s}$ \cite{2022NatAs...6.1085S} to the relative orbital velocity of VFTS 243 $v_{\rm{rel}}=\sqrt{G(M_{*}+M_{\rm{BH}})/a}\approx\sqrt{G(25+10\ M_{\odot})/(0.45\ \rm{AU}})\approx 260\ \rm{km/s}$, where $G$ is the gravitational constant, $M_{*}$ is the mass of the star, $M_{\rm{BH}}$ is the mass of the BH and $a$ is the semi-major axis. We conclude that for VFTS 243 the binary will likely widen due to rapid (``Jeans mode'') mass loss but the eccentricity will remain unchanged \cite{2021arXiv210709675S}. Finally, tidal dissipation drives the binary toward synchronization and circularization. The dynamical tide is the dominant tidal dissipation mechanism for main sequence O-type stars \cite{1975A&A....41..329Z}. For VFTS 243, the dynamical tide is not an efficient dissipation mechanism. The orbital period is too long ($\sim10$ d), the cirularization timescale is too long ($\gtrsim10^4$ Myr) and the lifetime of the system is too short ($\lesssim10$ Myr). Moreover, VFTS 243 is not yet synchronously rotating, which strongly suggest that tides have not played a significant role in the evolution of the system \cite{2022NatAs...6.1085S}.

\begin{figure}
    \centering
    \includegraphics[trim=1cm 7.5cm 0.8cm 8.cm, clip,width=0.48\columnwidth]{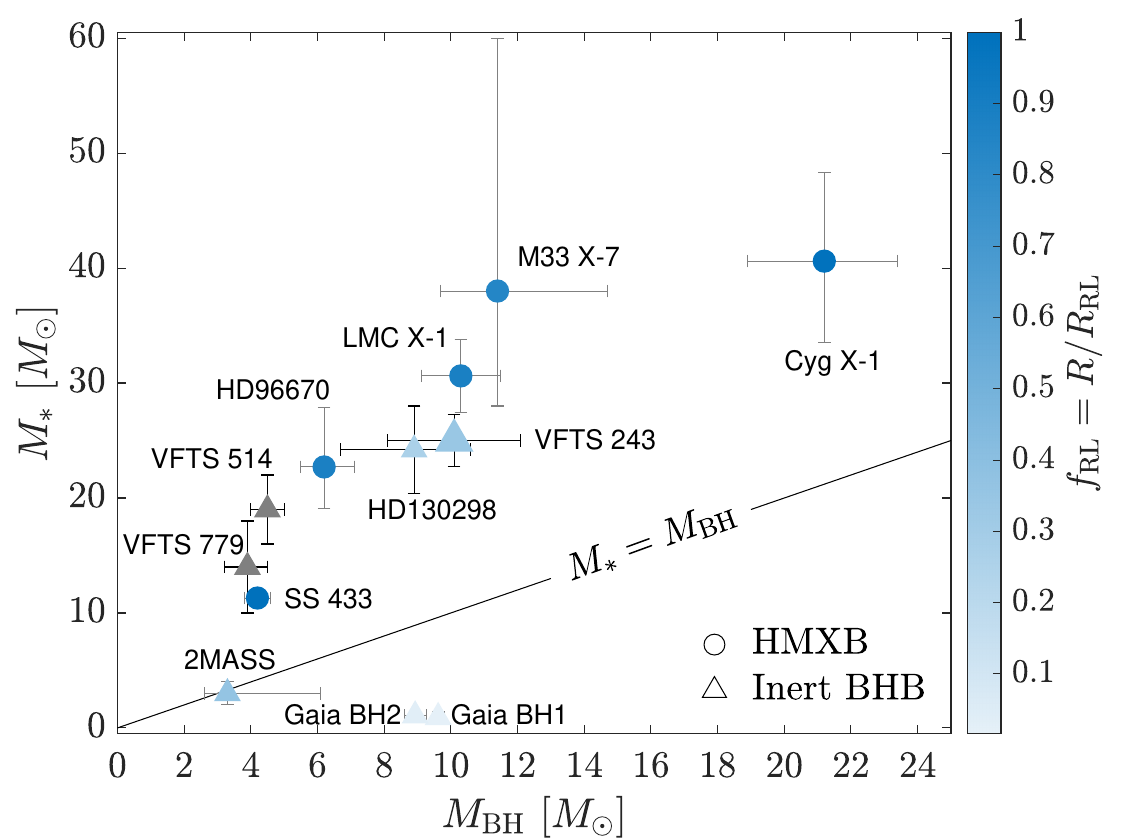}
    \includegraphics[trim=1cm 7.5cm 0.8cm 8.0cm, clip,width=0.48\columnwidth]{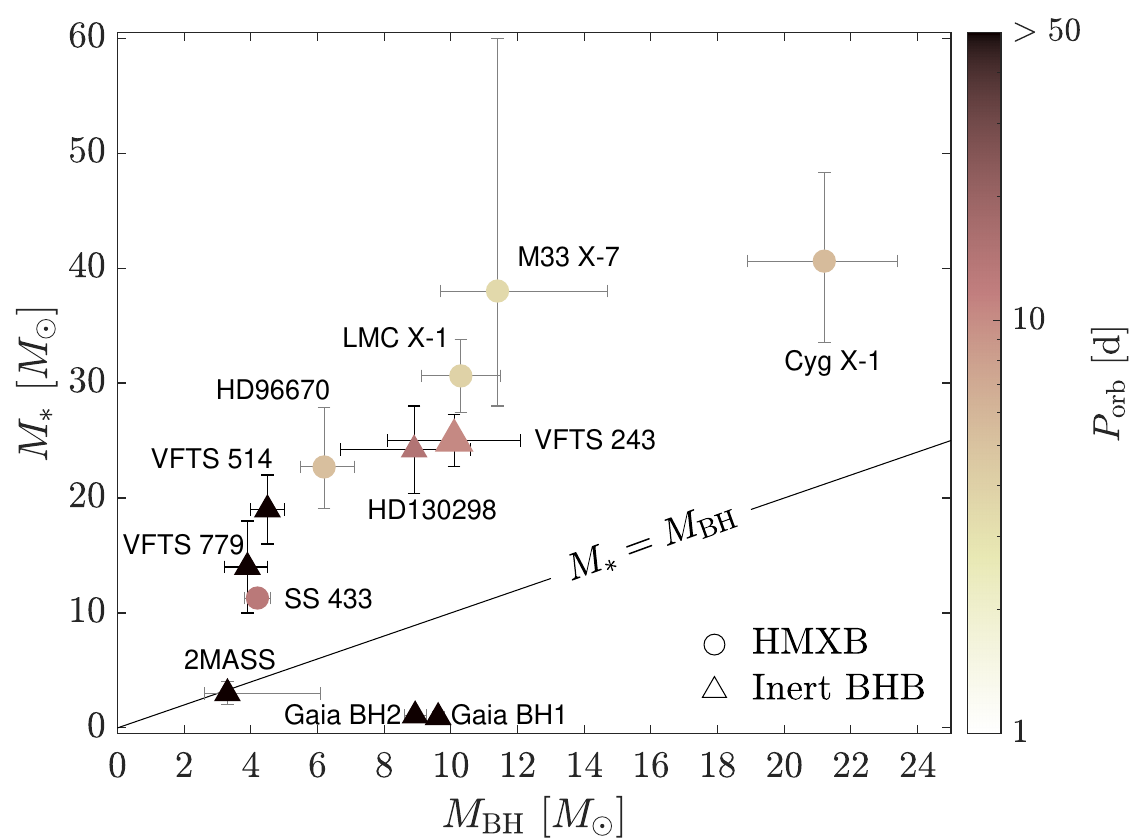}
    \caption{
    Additional orbital parameters of our observed sample of BH binaries with massive companions (cf. Fig.~\ref{fig:black-hole-binaries}). For each binary, we show the BH mass ($M_{\rm{BH}}$) on the abscissa, the stellar companion mass ($M_{\rm{*}}$) on the ordinate, and the Roche-lobe filling factor ($f_{\rm{RL}}$, left) or the orbital period ($P_{\rm{orb}}$, right) through the colorbar. We use circles to indicate high-mass X-ray binaries (HMXBs) and triangles for inert black-hole binaries (Inert BHB).
    The size of the markers is arbitrary, but the symbol of VFTS 243 is slightly larger for clarity. The radii of VFTS 514 and  VFTS 779 are not yet available and therefore the symbols are shown in gray in the left panel. For 2MASS J05215658+4359220 (2MASS), we use the lower limits on radius as reported in Ref.~\cite{2019Sci...366..637T}; this results in a lower limit for the Roche-filling factor.
    }
    \label{fig:more-orbital-properties}
\end{figure}

Finally, we comment on the main differences between the population of inert BH binaries and BH HMXBs \cite{2023A&A...677A.134N,2023A&A...671A.149F} (Fig.~\ref{fig:more-orbital-properties}). Most BH HMXBs are at orbital periods $\lesssim 6$ d; exceptionally, SS 433 has an orbital period of $13$~d and a circumbinary disk \cite{2010A&A...521A..81B,2019A&A...623A..47W}, a unique feature with respect to the other BH binaries in the sample. Excluding HD\,96670, for which the presence of a  BH still requires validation, SS\,433 is the only BH-HMXB with a massive companion for which an eccentric orbit has been derived. However, unlike for other HMXBs, the orbit of SS\,433 relies on radial-velocity measurements of variable emission lines, deeming its orbital solution, and in particular the eccentricity, susceptible to systematic errors. By contrast, inert BH binaries have orbital period $\gtrsim 10$ d.
For BH HMXBs, their short orbital period implies that tidal dissipation is more efficient than for wide inert BH binaries. The formation of an accretion disk around the BH requires some mass and angular momentum transfer from the star on the compact object; this process is not particularly efficient at circularizing the orbit but can in principle also change the eccentricity of the system \cite{2002MNRAS.329..897H,2012ApJ...747..111W}. Ref.~\cite{2012ApJ...747..111W} explored the natal kick of Cygnus X-1, as well as the orbital evolution of the system due to gravitational waves, stellar winds and tides. They find that stellar tides (winds) contribute the most to the change in the eccentricity (orbital period).

\end{document}